# Discovery of an Electron Gyroradius Scale Current Layer: Its Relevance to Magnetic Fusion Energy, Earth's Magnetosphere and Sunspots

**Authors:** Jaeyoung Park[1], Giovanni Lapenta[2], Diego Gonzalez-Herrero[2] and Nicholas Krall[1]

**Affiliations:**
1. Energy Matter Conversion Corporation, San Diego, CA, USA
2. Department of Mathematics, KU Leuven, University of Leuven, Belgium

**Abstract**

In the Earth's magnetosphere, sunspots and magnetic cusp fusion devices, the boundary between the plasma and the magnetic field is marked by a diamagnetic current layer with a rapid change in plasma pressure and magnetic field strength. First principles numerical simulations were conducted to investigate this boundary layer with a spatial resolution beyond electron gyroradius while incorporating a global equilibrium structure. The boundary layer thickness is discovered to be on the order of electron gyroradius scale due to a self-consistent electric field suppressing ion gyromotion at the boundary. Formed at the scale of the electron gyroradius, the electric field plays a critical role in determining equilibrium structure and plasma transport. The discovery highlights the necessity to incorporate electron gyroradius scale physics in studies aimed at advancing our understanding of fusion devices, the magnetosphere and sunspots.

**One Sentence Summary:** The thickness of the diamagnetic current layer between a plasma and a magnetic field is discovered to be on the order of electron gyroradius scale.

**Main Text:**

In many plasma systems, the plasma is surrounded by magnetic fields leading to a fascinating array of natural and manmade phenomena. Plasma jet formation from accretion disks, Earth's magnetosphere, sunspots and magnetic fusion devices are examples of plasma interaction with magnetic fields. At the boundary between the plasma and the magnetic field, if there is a change in plasma pressure or magnetic field strength, gyromotions of electrons and ions generate current, known as diamagnetic current, separating the plasma and magnetic field [1]. Among examples of plasma diamagnetic effects are the magnetopause in the Earth's magnetosphere, sharp boundary layers in magnetic cusp fusion devices, and the dark patches of sunspots [2-8]. In these systems, a diamagnetic current layer marks the boundary across which plasma penetration or loss to the magnetic field region is greatly reduced. The diamagnetic effect in these systems has been studied extensively, leading to the development of magnetohydrodynamics (MHD), the standard model for many solar, astrophysics and fusion plasmas over the past 50 years [9,10].

However, an *ab initio* solution of plasma diamagnetic effects had remained elusive with some of the most fundamental questions yet to be answered [11]. For example, there has been no definitive answer to the thickness of the diamagnetic current layer. Also unknown are the respective contributions of ions and electrons to the plasma diamagnetic current since there is significant difference in their gyroradii, a factor of 43 in the case of hydrogen ions at the same temperature as electrons. The lack of understanding remains because we are still trying to understand plasma dynamics at the scale of the electron the gyroradius, the fundamental, yet smallest, length scale of plasma diamagnetism. While there have been many theoretical and numerical studies to investigate the boundary layer structure, these studies have been limited due to geometrical complexities and technical challenges and have been unable to resolve electron gyroradius scale physics while incorporating the global equilibrium structure [12-16]. At the same time, a number of observations indicate the importance of electron scale phenomena at the boundary such as formation of electron scale ion flow in laboratory magnetic cusp experiments [17]. The recently launched Magnetospheric Multiscale (MMS) mission, designed to make electron scale plasma measurements, has started to generate observational data in the magnetopause demonstrating the importance of electron dynamics in magnetic reconnection and turbulence [18-21].

To explore the diamagnetic current layer on the electron gyroradius scale, we utilized a first-principles particle-in-cell (PIC) code, called the Energy Conserving semi-implicit model (ECsim), using its cylindrical coordinate implementation [22-24]. The ECsim simulates a collisionless plasma by solving Newton's equation for particle motion and Maxwell's equations for electric and magnetic fields, while conserving system energy. The simulations were conducted for a cylindrically symmetric magnetic cusp system known as the "Picket Fence" that was proposed as a magnetic confinement system for fusion energy, as shown in Figure S1 [25]. The magnetic field configuration of the picket fence is topologically identical to the dayside Earth magnetosphere, with the convex curvature of the Earth's dipole magnetic field facing the solar wind as well as the magnetic field in sunspots [6, 15, 26]. PIC simulations were conducted to investigate the boundary layer as a function of plasma pressure and ion mass with a spatial resolution beyond electron gyroradius while incorporating the global equilibrium structure. The exploration led to the discovery of a localized electric field at the electron gyroradius scale that transforms our understanding of plasma diamagnetic effects. Further details of the ECsim code and simulation method are given in the Method section in supplemental materials including Table ST1 summarizing the unit conversion between simulations and physical systems and Table ST2 summarizing simulation parameters used in the present study.

**Results**

**Steady-state equilibrium**

Figure 1 shows the steady-state equilibrium profile of a magnetic picket fence with a sharp boundary between plasma and magnetic field from Run 1 in Table ST2. From right to left, the magnetic field exhibits rapid decay across the boundary, leading to a field-free plasma region in the picket fence, as shown in Fig. 1(a). From left to right, the electron density profile exhibits similarly rapid decay across the boundary, leading to a plasma–free magnetic field region near the magnetic coils, as shown in Fig. 1(b). Across the boundary, layers of highly localized current are formed from plasma gyromotion separating plasma and magnetic field, as shown in Fig. 1(c). In addition, collimated ion flows are formed in the funnel-shaped cusp region, resulting in plasma leakage via the gap between the opposing sign of current layers, as shown in Fig. 1(d). For further analysis, three lines of interest are defined in Fig. 1(d) to describe the boundary between plasma and magnetic field, as described in the Method section. While exhibiting distinctively different equilibrium properties along Lines 1, 2 and 3, the different regions of equilibrium are interconnected by plasma motion and magnetic field, highlighting the necessity of incorporating the global equilibrium structure when investigating boundary layers.

Figure 2 shows steady-state equilibrium profiles as a function of volume averaged plasma pressure for Runs 2, 3, 4 and 5 in Table ST2 to investigate the change in equilibrium from plasma pressure change. The top row shows magnitude of magnetic field contours with magnetic field lines drawn to highlight the change in boundary location. The second row shows plasma diamagnetic current density, with the third row showing electron density and the fourth row showing ion mass flow in a radial direction. Along Line 1 as defined in Fig. 1(d), the boundary between plasma and magnetic field exhibits similar behavior for all four values of pressure. The increase in plasma pressure is balanced by the compression of the magnetic field. The boundary, marked by localized current layers moves toward the higher magnetic field region near the coils and the thickness of the current layer decreases. In comparison, there are significant differences in equilibrium along Lines 2 and 3. For pressures of $1.2 \times 10^{-6}$, $7.7 \times 10^{-6}$ and $5.2 \times 10^{-5}$ in normalized code unit (NCU), the plasma is still bounded by the magnetic wall of the picket fence. At these pressures, diamagnetic current layers converge toward narrow gaps in the cusp region coinciding with collimated ion flow. When the pressure is increased to $7.3 \times 10^{-5}$ in NCU, the magnetic wall fails along Line 2. While the current layers still converge toward each other, the gap between them is no longer narrow, with significantly wider density profile along Line 3 and increased radially outward ion flow.

Several features of the steady-state equilibrium in a magnetic picket fence in Figure 1 and 2 can be explained in a gross way with the standard MHD model. Eq. 1 shows the Momentum transport equation of the standard MHD model [1].

$$\rho \left( \frac{\partial V}{\partial t} + V \cdot \nabla V \right) = \frac{J \times B}{c} - \nabla p \quad (1),$$

where $\rho$ is the mass density of plasma, V is the plasma flow velocity, J is the current density, B is the magnetic field strength, c is the speed of light and p is the plasma pressure. In a steady-state equilibrium, the first term on the left-hand side (*lhs*) becomes zero, leading to the relationship known as the pressure balance equation among plasma flow, current, magnetic field and plasma pressure.

Along Line 1, the pressure balance between the plasma and magnetic field forms the boundary with diamagnetic current layers to match the change in magnetic field without plasma flow. With increasing plasma pressure, the boundary moves to the higher B-field region as the plasma works against the magnetic field that is compressible. Since the boundary layer thickness depends on the gyromotion of the plasma, the layer thickness decreases with increasing plasma pressure as previously discussed. In comparison, the pressure balance along Line 2 is different since the direction of the pressure gradient is along the direction of the magnetic field and the magnetic field can only exert forces on plasma that is moving across the field. The plasma pressure is instead balanced by the formation of radially outward plasma flow, as shown in Figures 1 and 2. Along Line 3, the magnetic field decreases in the plasma region with increasing plasma pressure due to the diamagnetic effects. If the plasma pressure becomes sufficiently high, the magnetic field inside the narrow gap becomes zero as the diamagnetic current provides complete shielding of the magnetic field by plasma. A further increase in the plasma pressure moves the boundary toward the coils similar to the boundary movement along Line 1, opening up the gap and leading to rapid leakage of plasma. Based on the similarity of magnetic field topology, the boundary layers along Line 1 and Line 3 correspond to the magnetopause and sunspot boundary, while collimated plasma flow along Line 2 corresponds to plasma loss to Earth's polar cusp region.

**Equilibrium as a function of ion mass**

Figure 3 shows plasma profiles in steady-state equilibrium for four different ion masses of $m_i = m_e$, 16 $m_e$, 1836 $m_e$ from Runs 6, 7, 8, and $m_i = 64$ $m_e$ from Run 4 in Table ST1. This study investigates the different roles of electrons and ions in determining equilibrium and boundary layer structure using a mass ratio between electrons and ions as a functional variable. The electron mass is kept constant and the ion mass is varied with the same temperature between electrons and ions. This results in an increase of ion gyroradius with respect to electron gyroradius. For example, the ion gyroradius for $m_i = 1836$ $m_e$ is 43 times larger than the ion gyroradius for $m_i = m_e$. The top row of Fig. 3 shows electron density profiles. The second row shows electron diamagnetic current density, and the third row shows ion diamagnetic current density. The radial ion flows are shown in the fourth row.

The key discovery illustrated in Fig. 3 is that the equilibrium profiles between the plasma and magnetic field remain nearly identical in their shape when the ion mass and ion gyroradius are varied by a factor of 1836 and a factor of 43 respectively. For example, the electron diamagnetic current layer occurs at the same location in space with only minor variation in its thickness. In terms of magnitude, electron density and electron diamagnetic current exhibit minimal change with ion mass variation. In contrast, there are significant decreases in ion diamagnetic current by a factor of 100 or more between $m_i = m_e$ and $m_i = 1836$ $m_e$, while the ion flow decreases by 8 times. These results were unexpected and prompted further investigation.

Figure 4 shows equilibrium profiles of radial electric field (top row) and axial electric field (bottom row) which can shed light on the unexpected discovery from Fig. 3 for the same set of runs. Along Line 1, there is little electric field in the case of $m_i = m_e$, consistent with the equal gyroradius of electrons and ions. In comparison, a localized electric field is formed and intensifies at the boundary with increase in ion mass to $m_i = 16$ $m_e$, 64 $m_e$ and 1836 $m_e$. The direction of the electric field is radially inward, thus in the direction of pushing ions from the boundary to the plasma region. With ions being pushed radially inward at the boundary, the electric field limits ion excursion into the magnetic field region, which in turn reduces the thickness of the boundary layer. The electric field also disrupts ion gyro-motion at the boundary leading to decreased ion contribution to the plasma diamagnetic current. In addition, the electric field intensity increases with ion mass in order to balance the larger ion gyroradius for heavier ions. While Line 1 is used to describe the role of the localized electric field, the presence of the electric field is seen on the entire surface of the boundary. By comparing the radial and axial electric field, it is shown that the direction of the electric field is normal to the magnetic field line and inward to the plasma region. As this localized electric field at the boundary could explain the unexpected discovery from Fig. 3, critical questions are the origin of the electric field and how to quantify its intensity.

**Analysis and Discussion**

To investigate the origin of the localized electric field, equilibrium profiles along Line 1 are examined in detail in Figure 5 for key plasma parameters in Eq. 1. In order to suppress numerical noise related to the use of discrete particles in the PIC simulation, the plot utilizes averaging of 20 steady-state plasma profiles, as discussed in the Method section. Figures 5(a), 5(b) and 5(c) show equilibrium profiles from Run 1 and Figure 5(d) shows ion diamagnetic current profiles as a function of grid resolution from Runs 1, 9, 10, 11 and 12 in Table ST2. As shown in Fig. 5(a), the boundary layer exhibits rapid change in plasma density, magnetic field and the diamagnetic current when the thickness of the current layer is ~0.6 as measured by full-width-half-maximum (FWHM). In addition, the radial profile reveals the occurrence of an electric field and its location with respect to the current layer. Figure 5(b) shows the electron and ion density profile in a semi-log plot, exhibiting exponential decay of both ion and electron density across the boundary layer. Figure 5(c) compares the radial electric field and the gradient of ion pressure divided by ion density showing that the electric field develops at the boundary as the ion pressure decreases. While detailed simulation parameters and results are summarized in Table ST1, some relevant values are given here for Run 1. In NCU, the simulation utilizes an electron thermal velocity of $7.35 \times 10^{-2}$ times the speed of light with an electron mass of 1/64 and an ion thermal velocity of $9.2 \times 10^{-3}$ times the speed of light with an ion mass of 1, with the speed of light and charge of electrons and ions normalized to 1. As shown in Fig. 5(a), a mean value of magnetic field magnitude is $4.1 \times 10^{-3}$ in the current layer. This leads to the thermal electron gyroradius of 0.28 and the ion gyroradius of 2.24 since the gyroradius is given as the thermal velocity multiplied by the particle mass and the inverse of magnetic field in NCU. Therefore, the current layer thickness of 0.6 corresponds to approximately twice the electron gyroradius and a quarter the ion gyroradius.

During the analysis to quantify the electric field intensity, we have also discovered the importance of spatial resolution for PIC simulation, as shown in Fig. 5(d). Here we conducted a series of convergence tests with respect to the grid resolution from g0 (45x30) to g1 (90x60), g2 (180x120), g3 (360x240) and g4 (540x360) corresponding to the grid size varying from 3.6, 1.8, 0.9, 0.45 and 0.30 times the electron gyroradius at the boundary. Fig. 5(d) shows ion diamagnetic current density as a function of grid resolution. The simulation reaches a converged solution for g3 and g4, while g2 results seem to be reasonably close to the converged solution with respect to ion diamagnetic current density. On the other hand, without sufficient grid resolution, such as in the g0 and the g1 cases, numerical inadequacy leads to over-estimation of ion diamagnetic current density and its layer thickness in the boundary layer.

The results shown in Figures 5(a), 5(b) and 5(c) are unexpected and outside the standard MHD model, whose solution of the current layer does not include the electric field. Instead, we compare the results with the equation known as generalized Ohm's law which relates the current to the electric field, as shown in Eq. 2 [27].

$$E = -\frac{V_i \times B}{c} + \frac{1}{ne} \times \frac{J \times B}{c} - \frac{\nabla p_e}{ne} \qquad (2),$$

where E is the electric field, $V_i$ is the velocity of ions, n is the plasma density, e is the electron charge and $p_e$ is the pressure of electrons. It is noted that time varying terms are ignored in Eq. 2 as we are interested in steady-state equilibrium. Electron inertia terms, the plasma resistivity term and other higher order terms such as off-diagonal pressure tensor terms are ignored as well as the difference between the electron density and the ion density. First, we note that the first term on the right-hand side (*rhs*) can be ignored at the boundary along Line 1 since there is no plasma mass flow as shown in Figures 1. We can then utilize Eq. 1 to replace the J x B term, the second term on the *rhs* with the total pressure gradient reducing Eq. 2 into a simple relation between the electric field and the ion pressure at the boundary.

$$E = \nabla p_i/ne \text{ or, } E = kT_i \nabla n_i/ne \qquad (3),$$

where $p_i$ is the pressure of ions, k is the Boltzmann coefficient, $T_i$ is the ion temperature and $n_i$ is the ion density.

This relationship between the electric field and the ion density gradient is the key discovery. Since the electric field intensity is proportional to the ion density gradient, it highlights the importance of fully resolving length scale down to the electron gyroradius in determining ion dynamics at the boundary. An approximate solution of this relationship can be expressed as $n_i \sim n_0 \exp(eE_0(r-r_b)/kT_i)$, where $n_0$ is the ion density at the boundary location at $r=r_b$ and $E_0$ is mean electric field value in the boundary layer. The observed exponential decay of ion density shown in Fig. 5(b) agrees with this solution. Finally, Fig. 5(c) shows an agreement between the radial electric field and the gradient of ion pressure divided by the ion density, as shown in Eq. 3.

To further understand the role of the electric field, Figure 6 compares single particle ion trajectories in the equilibrium from Run 1 with and without the electric field along Lines 1 and 2 with green dots representing origins of their trajectories. All ions begin their motion with the same velocity vector angled at 15 degrees between radial velocity and axial velocity, and their kinetic energy equal to twice the kinetic energy of plasma injection during the initialization phase. As shown in Figures 6(a) and 6(b), ion motions exhibit sharp reflection at the boundary due to the presence of the electric field. In comparison, ions would penetrate significantly deeper across the boundary layer if the electric field is ignored, as shown in Figures 6(c) and 6(d). Along Line 1, the sharp reflection of ions at the boundary is consistent with the exponential ion density decay, with the characteristic decay length comparable to electron gyroradius. The electric field also contributes to the ion flow collimation along Lines 2 and 3, with a width of ion flow significantly less than ion gyroradius, while suppressing plasma leakage, as shown in Figure 6(b). Without the electric field at the boundary, the width of the ion flow would be significantly wider, as shown in Figure 6 (d).

The results from Fig. 6 show that the main role of the electric field at the boundary is to limit ion excursion at the boundary, which in turn limit charge separation between electrons and ions, as shown in Fig. 5(b). As the ion excursion is suppressed at the boundary, the ion density decreases rapidly at the boundary. This leads to the decrease of ion diamagnetic current with the ion diamagnetic current layer thickness comparable to the electron diamagnetic current layer thickness, as shown in Fig. 3. At the same time, the gradient of ion density or ion pressure term becomes significant, which gives a rise to the electric field at the boundary as shown in Fig. 5(c) and Eq. 3. Therefore, this electric field can be described as the self-consistent field since it occurs to prevent additional charge separation beyond the generation of the electric field leading to the electron gyroradius scale boundary layer. Separately, Fig. 6 also provides a clear explanation for the previously unresolved rapid formation of collimated ion flow observed since the ion collimation is caused by the self-consistent electric field rather than ion gyromotion [17].

Our central result is that diamagnetic effects of plasma can produce electron-scale boundary layers across which current, density and magnetic field exhibit sharp transition on electron gyroradius scale length. This discovery comes at a fortuitous moment when the recently launched Magnetospheric Multiscale (MMS) mission has the capability to capture electron scale plasma dynamics both as a function of time from the high cadence of its instrumentation and space because of the short distance between its four spacecraft. It should therefore be possible in principle to observe our predicted structures. For example, Burch et al reported an observation of electron scale current layers in the electron diffusion region of magnetic reconnection sites during magnetopause crossings by MMS spacecraft and identified the critical role of electron dynamics in triggering magnetic reconnection [18]. Electron scale current layers have also been observed in the magnetosheath as part of the turbulent cascade with the observation of the electron jets in the absence of ion reconnection signature. [19]. In addition, small scale magnetic holes produced by diamagnetic effects have been observed where the magnetic hole exhibits electric and magnetic field boundary structures on the order of ~30 km compared to the ion gyroradius of 100 – 1,000 km [20]. Finally, the electron scale diamagnetic current layer has also been observed with the current produced predominantly by the divergence of pressure tensor near a magnetic reconnection region [21]. While exact mechanisms producing such electron scale current layers and field structure requires further investigations, the electron scale diamagnetic current layer discovered in our simulation could be a possible source of these electron scale plasma structures.

**Summary and Conclusion**

The fully kinetic first principles simulation resolving electron gyroradius scale length reported here led to the discovery of a localized and self-consistent electric field that plays a critical role in the boundary layer marked by a diamagnetic current between the plasma and surrounding magnetic fields. This electric field arises from the ion density or pressure gradient at the boundary and its main role is to limit charge separation between electrons and ions. By suppressing ion excursion across the boundary, the electric field leads the current layer thickness to the length scale of the electron gyroradius, the smallest and most fundamental length scale in the magnetic properties of plasma, instead of the much larger ion gyroradius. The electric field also affects plasma transport across the boundary by collimating plasma flow in the cusp region flow and reducing plasma leakage.

The discovery of this localized electric field highlights the necessity to incorporate electron gyroradius scale physics in future studies aimed at advancing our understanding of fusion device performance, the magnetosphere and sunspots. In the case of magnetic cusp fusion devices, the discovery encourages the resumption of research into magnetic cusp devices as potential thermonuclear fusion energy reactors. Magnetic cusp systems, in addition to their proven plasma stability and engineering simplicity, are one of the few magnetic fusion devices that allow direct injection of a charged particle beam into the central region [28,29]. The use of an electron beam may allow control of the electric field at the boundary toward the further improvement of plasma confinement in conjunction with flow collimation [30]. A numerical capability to accurately calculate the electric field offers the tantalizing potential to improve the performance of magnetic cusp devices toward net fusion energy generation. While the present work is focused on systems where the diamagnetic current layer separates a field-free plasma and a plasma-free magnetic field, the localized electric field may also play a role in plasma equilibrium and confinement at the boundaries of other fusion devices, such as the tokamak, stellarator, magnetic mirror, and Field Reversed Configuration (FRC). This is because the boundary layers of these devices are subject to the same set of equations such as the pressure balance equation and the generalized Ohm's law. In the case of the Earth's magnetosphere, incorporating an electron gyroradius scale boundary layer in the quintessential equilibrium between the solar wind plasma and the Earth's magnetic field will provide new insights into magnetic reconnection and plasma turbulence. This is because the gradient scales of the current layer and plasma pressure play a critical role in the reconnection rate and turbulence spectrum in magnetic reconnection and plasma turbulence. Extending experimental and theoretical tools toward electron gyroradius scale phenomena will help to take full advantage of the recently launched MMS mission.

**Acknowledgments:**

The authors would like to thank Prof. John F. Santarius at Univ. Wisconsin, Madison and Prof. Yong Seok Hwang at Seoul National Univ. of Korea for their helpful discussion on plasma diamagnetic effects and Dr. Alan Roberts at Energy Matter Conversion Corporation and Mr. John Draper at Khon Kaen Univ. of Thailand for their assistance in editing of the manuscript. The initial funding for this work came from the internal corporate research and development funds of Energy Matter Conversion Corporation. This research used computational resources provided by NASA NAS and NCCS High Performance Computing, by Flemish Supercomputing Center (VSC), and by PRACE Tier-0 allocations. Park contributed to the conceptualization of the problem, conducted simulations and analysis, and prepared the manuscript. Lapenta contributed to the development of the simulation code, conducted the analysis, and edited the manuscript. Gonzalez-Herrero contributed to the development of the simulation code and provided the support for conducing simulations. Krall contributed to the conceptualization of the problem and supported the analysis.

**List of Supplementary Materials:**

- Materials and Methods.
- Table ST1-ST2.
- Figures S1-S5.
- References (31,32)
- Movies M1-M2.

**Supplementary Materials:**

**Materials and Methods**

**a) Magnetic picket fence system**

Figure S1 shows a schematic of a magnetic picket fence system used in the simulation. It consists of series of circular coils arranged along the vertical axis with opposite coil current direction between adjacent coils. These coils produce zero magnetic field near the central region near the axis and form a magnetic field wall near the coils. The magnetic picket fence was proposed by Tuck in 1954 as a magnetic confinement system to produce thermonuclear fusion reactions [25]. The picket fence is one of the magnetic confinement systems called "magnetic cusps", that are known to be stable against many of plasma instabilities [Krall, Berkowitz]. In this report, the magnetic picket fence system was chosen for the following reasons.

1. Perfect magnetic field shielding has been experimentally observed in various magnetic cusp systems designed for fusion energy research [6,14,31,32]. As such, the magnetic cusp system is well suited to investigate diamagnetic effects of plasma. In addition, the magnetic field configuration of the picket fence is topologically identical to the dayside Earth magnetosphere with the convex curvature of the Earth's magnetic field facing the solar wind as well as the magnetic fields of sunspots [6,15,26].

2. Due to their favorable magnetic field curvature, magnetic cusp systems have been shown to be stable against most, if not all, of macroscopic plasma instabilities in theory. This is because plasma must do work compressing the magnetic field if it expands at the boundary since the magnetic field is curved into the plasma on every surface. The lack of plasma instabilities in magnetic cusp systems has also been reported in many past experiments. This allows the simulation of underlying equilibrium to reach steady-state or at least quasi-steady state in a couple of plasma transit times, as determined by the slower species, i.e. ions.

3. A magnetic picket fence can be simulated with the periodic boundary condition in the axial direction. Most proposed fusion reactor configurations based on magnetic cusp system utilize many pairs of magnetic coils to provide sufficient reactor volume and needed confinement [30]. In the case of a magnetic picket fence utilizing many pairs of coils along the axial direction, the periodic boundary condition is a good approximation in the central region of the picket fence as shown in Fig. S1. In the present study, a set of 27 coils are used in the simulation to provide the external magnetic field that is periodic along the symmetric axis with 5 coils in the middle being shown in Fig. S1. In addition, the plasma refueling can be achieved by injection from the both ends to achieve steady-state operation, corresponding to volumetric plasma injection near the axis used as in the simulation.

In summary, plasma dynamics in the magnetic picket fence system has been simulated using a fully kinetic PIC code to investigate diamagnetic effects. The simulations utilize the cylindrical symmetry in the angular direction and the periodic boundary condition in axial direction while preserving a dipole nature of the magnetic field in the simulation. A steady-state equilibrium is produced by injection of the plasma in the central part of the picket fence and the plasma loss boundary that absorbs ions and electrons that leak out of the picket fence system, as shown in Fig. S1. It is noted that fully a kinetic PIC simulation of the diamagnetic current layer requires significant High Performance Computing (HPC) resources even in the simple geometrical setup of an axisymmetric magnetic picket fence system. Typical runs employ between 300 and 1200 CPUs and require between 10,000 and 150,000 CPU hours to simulate the steady-state equilibrium while resolving electron gyroradius with satisfactory numerical convergence.

**b) Code Description**

For the study reported, we utilize a fully kinetic description of the equilibrium between plasma and magnetic field, where both electrons and ions are followed as particles interacting via electric and magnetic fields generated by the particles themselves as well as by the coils. The approached followed is the electromagnetic particle in cell (PIC) method. The full set of Maxwell's equations is discretized on a grid where particle moments are collected via first order basis spline interpolation to calculate the sources for Maxwell's equations. In the present paper, we utilized the Energy Conserving semi-implicit method (ECsim) in its cylindrical implementation called ECsim-CYL [22-24] based on azimuthal symmetry. The ECsim-CYL solves the field equations in two-dimensional (2D) cylindrical coordinates using a finite volume method. For the particles, it solves all three components of velocity vectors, while only keeping radial and axial coordinates of particle positions. The numerical algorithm of ECsim-CYL has been tested previously for accuracy and convergence [24].

We utilized the ECsim-CYL code to investigate the plasma diamagnetic effects for the following reasons. The ECsim-CYL conserves the system energy precisely down to machine precision even when the grid and time resolution severely under-resolve the electron plasma frequency or the electron Debye length. This energy conservation allows the simulation to operate without any artificial smoothing. While the field or the particle moment smoothing helps with noise and numerical stability, the use of smoothing leads to the violation of energy conservation and disrupts the diamagnetic boundary layer leading to an artificially greater layer thickness caused by numerical effects rather than physical effects. Though, in principle, it is possible to avoid the under sampling of electron plasma frequency or Debye length, the numerical cost is very high, about a factor of 100 or more for the plasma parameter spaces as shown in Table ST1. This is because the Debye length is about a factor of 10 smaller than the electron gyroradius. This additional computational cost needs to be multiplied by each dimension, leading to a factor of 100 increase in 2D cylindrical geometry. On the other hand, the implicit PIC codes, such as ECsim, have successfully demonstrated the ability to resolve critical electric field generation regarding charge separation between electrons and ions even when they are under sampling the Debye length [24]. Considering that each run in Table ST2 already requires 10,000 to 150,000 CPU hours to generate an equilibrium solution, the use of the energy conserving algorithm of ECsim was critical to resolve electron gyroradius scale physics in the boundary layer with built-in energy conservation. It is noted that the system energy is conserved to machine precision at all resolutions reported.

The ECsim code uses normalized code units (NCU) that are non-dimensional. The use of NCU allows the simulation results to be converted to various physical systems over a wide range of parameters. Two examples are shown in Table ST1 where the simulation results are converted to plasma parameters relevant to magnetic fusion devices and the Earth's magnetopause. In ECsim, the density is normalized to a reference density $n_0$. Time is normalized to the ion plasma frequency, $\omega_{pi}$, determined by the reference density, as $\omega_{pi} = (n_0/m_i)^{0.5}$, where $\omega_{pi}$ is the ion plasma frequency and $m_i$ is the ion mass in NCU. Electron plasma frequency is defined similarly, as $\omega_{pe} = (n_0/m_e)^{0.5}$, where $\omega_{pe}$ is the electron plasma frequency and $m_e$ is the ion mass in NCU.

Velocities are normalized to the speed of light that is set at 1. Distances are normalized to the ion inertial length, as $d_i = c/\omega_{pi}$. Magnetic fields are normalized to satisfy the following relation regarding the plasma gyroradius, $r_{e,i} = m_{e,i} v_{e,i}^{th}/B$ in NCU, where $r_{e,i}$ is the electron and ion gyroradius, $v_{e,i}^{th}$ is the electron and ion thermal velocity and B is the magnetic field magnitude. This normalization allows the conversion of simulation results to different plasma temperatures by simultaneously varying the thermal velocity and the B-field value while maintaining the same plasma gyroradius. Separately, the charge of electrons and ions is set at 1 and the permittivity and permeability of free space, and Boltzmann coefficient are also set at 1 in NCU. Here are the formulas to convert values in NCU to physical units once the reference density, $n_0$ is chosen along with the plasma temperature or the reference velocity. The magnetic field $B^*$ in physical unit is given as $B^* e^*/(m_i^* \omega_{pi}^*) = m_i^{0.5} B$, where the left-hand side (*lhs*) quantities are in SI unit with $B^*$ in Tesla, $e^*$ is the electron charge in Coulomb, $m_i^*$ is the ion mass in kg and $\omega_{pi}^*$ is ion plasma frequency in rad/s, while the right side quantities are in NCU where B is the magnetic field magnitude and $m_i$ is the ion mass used in the simulation. The electric field $E^*$ in physical unit is given as $E^* e^*/(c^* m_i^* \omega_{pi}^*) = m_i^{0.5} E$, where the *lhs* quantities are in SI unit with $E^*$ in V/m, $c^*$ is the speed of light at $3 \times 10^8$ m/s, while the right-hand side (*rhs*) quantities are in NCU where E is the electric field magnitude in the simulation. The current density $J^*$ in physical unit is given as $J^* e^* \mu_0^* d_i^*/(c^* m_i^* \omega_{pi}^*) = m_i^{0.5} J$, where the *lhs* quantities are in SI unit with $J^*$ in A/m², $\mu_0^*$ is permeability of free space at $4\pi \times 10^{-7}$ Hm⁻¹, and $d_i^*$ is the ion inertial length in meters, while the right side quantities are in NCU where J is the current density in the simulation.

### c) Simulation setup

Each simulation begins with plasma injection of electrons and ions from the center of the picket fence to achieve the preset plasma pressure and ends when the equilibrium reaches quasi-steady state as shown in Figure S2. The red graded region in Figure S1 shows the area of volumetric plasma injection. In NCU with the length scale normalized to ion inertial length, the size of the simulation domain is 45 in radius and 30 in height or axial length, as shown in Figure 1(d), while the injection region is 9 in radius and 30 in height. Coils in the picket fence have a diameter of 60 and the spacing between two adjacent coils is 15. During initialization, ions and electrons are injected with the same temperature with an electron thermal velocity of $7.35 \times 10^{-2}$ times the speed of light in NCU. The ion thermal velocity, on the other hand, is adjusted as a function of ion mass to maintain the same temperature for both species. A typical time step is $0.25/w_{pi}$, during which a thermal electron travels $1.84 \times 10^{-2}/d_i$ and a thermal ion travels $2.3 \times 10^{-3}/d_i$ in NCU. Once injected, the plasma expands and fills the picket fence system while interacting with the externally applied magnetic field. During expansion, the plasma expels a magnetic field from the plasma and forms a boundary. The temporal duration of the injection phase is $8,000/w_{pi}$, corresponding to 10 times the electron transit time or 1.2 times the ion transit time for the ion mass of $m_i = 64\, m_e$. The transit time is defined as the time for thermal ions and electrons to move across one coil diameter. The injection is conducted incrementally for 160 times during the initialization phase with an equal amount of plasma injections leading to gradual increases in the total kinetic energy of the plasma in the picket fence and plasma diamagnetic effects as shown in the top row of Fig. S2. Incremental injection is used to build up plasma pressure in the picket fence gradually without generating shocks or significant plasma flow, to investigate the quiescent equilibrium between the static plasma pressure and magnetic field pressure.

Once the preset plasma pressure is reached in the picket fence, the initialization phase is complete and the system is relaxed toward a steady-state, as shown in Fig. S2. During the steady-state phase, plasma is maintained by incremental injection in the same central region of the picket fence to replenish the loss of plasma from the picket fence to the loss boundary at the right end of the simulation domain. The loss boundary is simulated as an absorbing wall for particles and electromagnetic waves, shown as a dotted line in Figure S1. It is located at r=42, away from the coils at r= 30 to prevent the presence of the wall from affecting plasma equilibrium inside the picket fence. Nominally, the injection rate to sustain the plasma during the steady-state phase is ~18 times lower than the injection rate during the initialization phase. For example, a charge injection rate of ~ 3 per $22.5/w_{pi}$ is utilized to maintain a constant total charge of $1.89 \times 10^4$ in the picket fence system for Run 1 in Table ST2 during the steady-state. This injection rate corresponds to a particle confinement time of $1.1 \times 10^5/w_{pi}$, equivalent to ~135 electron transit time or ~17 ion transit time. While the injection rates for ions and electrons are allowed to vary from each other while replenishing their respective charge loss, the plasma loss quickly satisfies the ambipolar condition with equal loss of electrons and ions from the picket fence to the absorbing wall as shown in the bottom row of Fig. S2. On the other hand, plasma injection during the steady-state phase requires more kinetic energy per injected particles compared to plasma injection during the initialization phase by a factor of 2.5 to 3. This is equivalent to the energy confinement time of the system being 2.5 to 3 times shorter than the particle confinement time. A shorter energy confinement time is typical in most plasma systems as higher energy particles leave the systems faster than lower energy particles.

Nominally, the simulation is conducted for a minimum of 2.2 times the ion transit times after the initialization phase to ensure steady-state. By then, all equilibrium properties such as plasma density, current density, plasma pressure, plasma flow and magnetic field are nearly constant in space and time. As shown in M1 and M2 (movies), the location and the width of the boundary layer are constant with less than one to two pixels variation. The M1 is from Run 8 in Table ST2 that covers 7 ion transit times from beginning to an end and the M2 is from Run 1 in Table ST2 that covers to 3.5 ion transit time from the beginning to an end. Note that the sudden changes in radial ion mass flow, shown in the movies, are related to transition from the initialization phase to the steady-state phase, which involves change in plasma injection rate by a factor of ~18. For the present study, we have conducted systematic studies of equilibrium between the plasma and magnetic field as a function of plasma pressure and ion mass for a constant electron mass. In addition, several additional tests were conducted to ensure the numerical convergence with variation in grid size, time step and number of simulation particles. Table ST2 summarizes the key parameters used in the simulations.

**d) Numerical Convergence**

To ensure the accuracy of the simulation results, we conducted various convergence tests consisting of changes to the number of particles per each grid, the size of the time step and grid resolution. The simulation parameters for these runs are shown in Table ST2 and the results of the convergence tests are shown in Figure S3 along Line 1. Fig. S3(a) shows the number of simulation particles in each grid across the entire length of Line 1. In comparison, Fig S3(b) shows the plasma diamagnetic current density at the boundary layer. Four different particle number settings are used representing the number of particles at the boundary region as noted in the legend. For example, the setting of N80 denotes the simulation where the number of plasma particles at the boundary region is ~80 and the setting of N5120 denotes the simulation where the number of plasma particles at the boundary region is ~5120. As shown in Fig. S3(b), the simulation reaches numerical convergence once there are sufficient plasma particles over 300 at the boundary region. In comparison, all runs discussed in the present study have between 1000 and 1600 particles at the boundary region. In Figures S3(c) and S3(d), electron and ion diamagnetic current density are shown at the boundary layer along Line 1 for three different time steps of 0.125, 0.25 and 0.5. The simulation reaches numerical convergence once a time step of 0.5 or smaller is used. All runs discussed in the present study utilize the time step of 0.25 with an exception of Run 1 where the time step of 0.125 is used in conjunction with the g4 (540x360) grid size. The need to use a smaller time step is the main numerical constraint of implicit particle codes such as ECsim to ensure that the particles do not travel more than one grid over a single time step. Figure S3(e) shows the radial electric field and Fig. S3(f) shows the electron diamagnetic current density as a function of grid resolution with the grid resolution varying from g0 (45x30) to g1 (90x60), g2 (180x120), g3 (360x240) and g4 (540x360), corresponding to the grid size varying from 3.6, 1.8, 0.9, 0.45 and 0.30 times the electron gyroradius at the boundary. As discussed in the main text related to Fig. 5(d), the simulation reaches a converged solution for g3 and g4, while g2 results seem to be reasonably close to the converged solution. On the other hand, without sufficient grid resolution, such as in the g0 and the g1 cases, it is difficult to identify and quantify the electric field while not being able to resolve the gradient scale length of boundary layer properly. The numerical inadequacy leads to under estimation of radial electric field, consistent with Eq. 3 in the main text where the electric field is expressed as the gradient of ion pressure term.

**e) Additional Notes**

As an addendum to Fig. 3, the study included additional ion masses of $m_i = 4\ m_e$ and $m_i = 256\ m_e$ and the results for the total diamagnetic current and ion diamagnetic current are shown in Figure S4. It is noted that the integral of the diamagnetic current gradually decreases with increasing ion mass with its value for $m_i = m_e$ being 25% higher than the value for $m_i = 1836\ m_e$, shown in Fig. S4(a). In comparison, the ion diamagnetic current decreases rapidly with increasing ion mass while the thickness of ion diamagnetic current layer varies, as shown in Fig. S4(b). Note that Fig. S4(b) utilizes the logarithmic scale for ion diamagnetic current density to illustrate rapid decreases of ion diamagnetic current with increasing ion mass.

Figure S5 compares the simulation noises for the radial electric field and electron diamagnetic current density along Line 1 at the boundary from Run 1. Due to substantial noise in the electric field related to the particle discreteness as shown in Fig. S5(a), the electric field results in Fig. 5(a) and Fig. 5(c) are presented with simple averaging from 20 different time cycles during the steady-state, shown as thick blue dotted line in Fig. S5(a). In comparison, other plasma parameters such as the electron diamagnetic current exhibit minimal noise during steady-state as shown in Fig. S5 (b), hence averaging is not needed.

While the investigation of plasma instability is outside the scope of the present study, we report the following observations from the simulation results. The boundary layer along Line 1 shows no sign of plasma instability as shown in Figures 1-4. This is consistent with the standard MHD stability model where the convex curvature of magnetic field facing plasma provides stabilizing force. In comparison, the boundary along Lines 2 and 3 exhibits signs of plasma instabilities in particular where the plasma flow is significant and the magnetic field curvature is reduced compared to Line 1. The sign of plasma instabilities is best shown in Fig. 4, where the electric field exhibits electron gyroradius scale structures in the funnel shaped cusp region even in the case of an equal mass of electrons and ions. One possible explanation is that this structure may be related to plasma instability from steep gradient in plasma current where the stabilizing effect of magnetic field line curvature is weak.

**Supplemental References**

31. R. E. Pechacek, J. R. Greig, D. W. Koopman, A. W. DeSilva, Measurement of the plasma width in a ring cusp, *Phys. Rev. Lett*. **45**, 256-259 (1980).

32. A. Kitsunezaki, M. Tanimoto, T. Sekiguchi, Cusp confinement of high beta plasmas produced by a laser pulse from a freely-falling deuterium ice pellet, *Phys. Fluids* **17**, 1895-1902 (1974).

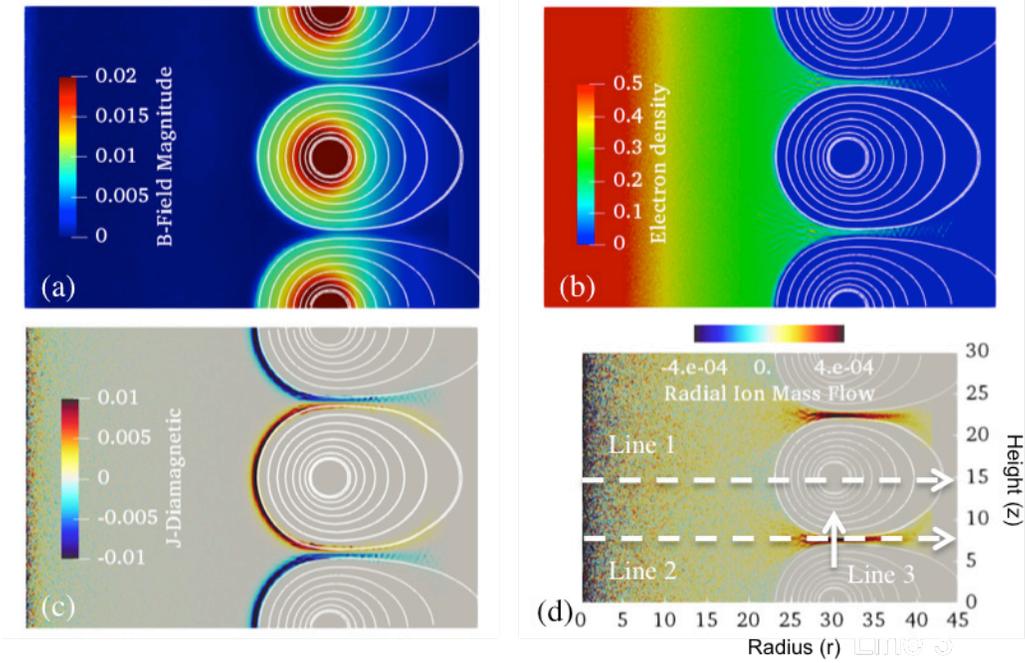

Figure 1. Steady-state equilibrium profiles. Equilibrium profiles show contours of (a) magnitude of B-field, (b) electron density, (c) diamagnetic current density and (d) radial ion mass flow with magnetic field lines from Run 1 in Table ST2. Three lines of interest are defined in Fig. 1(d) for further analysis with Line 1 (r=0 to r=45 at z=15), Line 2 (r=0 to r=45 at z=7.5) and Line 3 (r=30, z=4 to z=11).

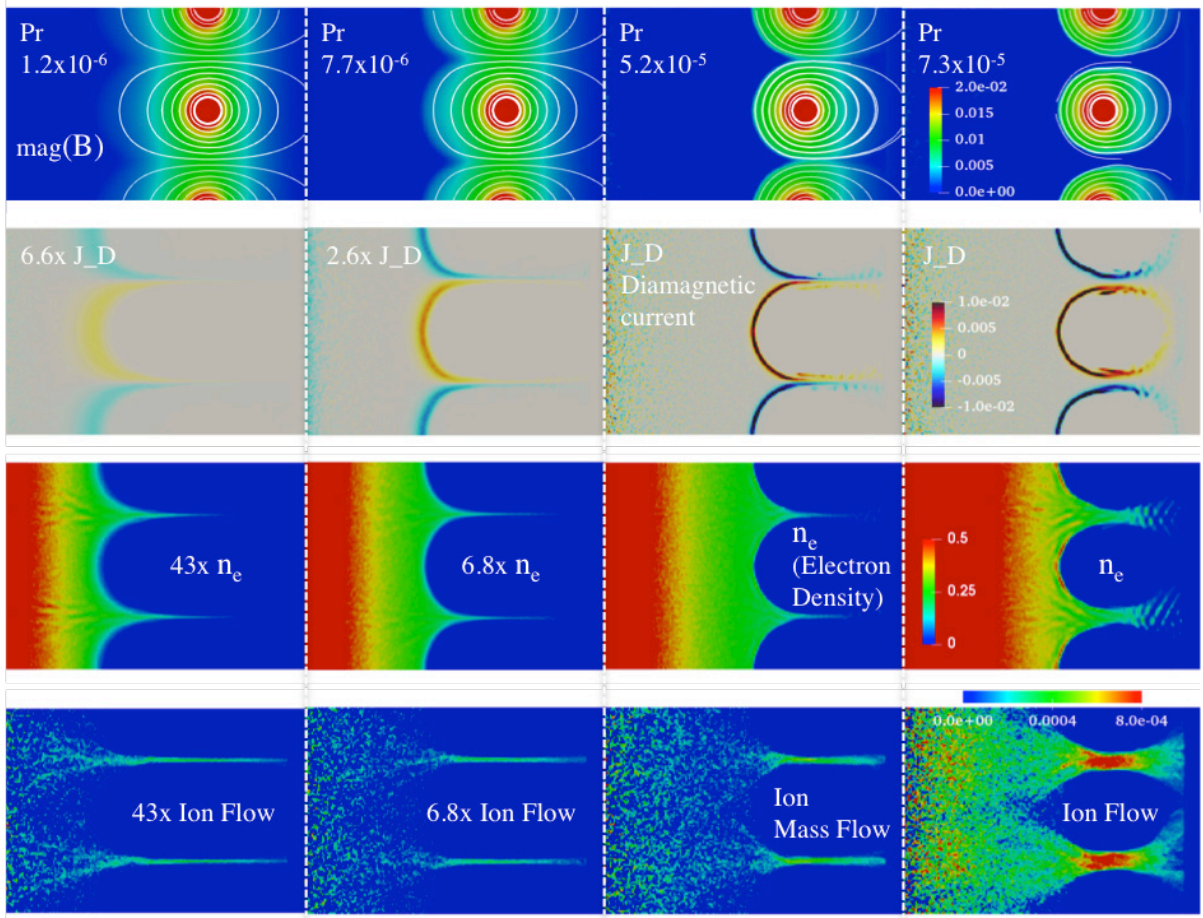

Figure 2. Steady-state equilibrium profiles as a function of plasma pressure. Equilibrium profiles show magnetic field magnitude (top row), diamagnetic current density (second row), electron density (third row), and radial ion mass flow as a function of plasma pressure (fourth row) in the picket fence for four different volume averaged pressure values of $1.2 \times 10^{-6}$ (First column from the left). $7.7 \times 10^{-6}$ (second column), $5.2 \times 10^{-5}$ (third column) and $7.3 \times 10^{-5}$ (fourth column) from Runs 2, 3, 4 and 5 in Table ST2. The current, density and flow profiles from the pressures of $7.7 \times 10^{-6}$ and $5.2 \times 10^{-5}$ are multiplied by different factors as noted in plots to utilize the same color scales shown on the fourth column.

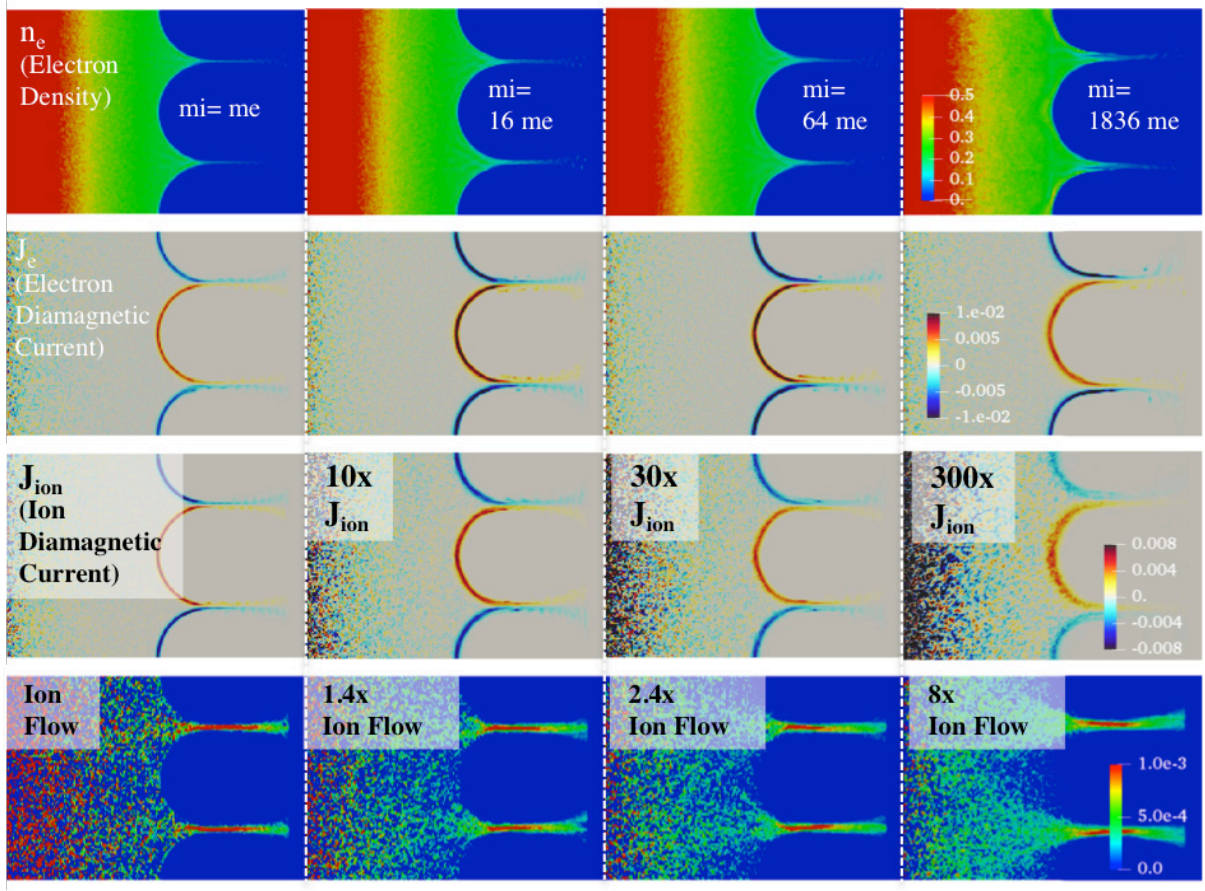

Figure 3. Steady-state equilibrium as a function of ion mass. Equilibrium profiles show electron density (top row), electron diamagnetic current density (second row), ion diamagnetic current density (third row) and radial ion mass flow (fourth row) for 4 different ion masses of $m_i = m_e$ (first column from the left, from Run 6), $m_i = 16\ m_e$ (second column, from Run 7), $m_i = 64\ m_e$ (third column, from Run 4) and $m_i = 1836\ m_e$ (fourth column, from Run 8). To fit in the same color scales shown on the fourth column, the ion diamagnetic current density and ion flow are multiplied by different factors as noted in plots to utilize the same color scales shown on the fourth column.

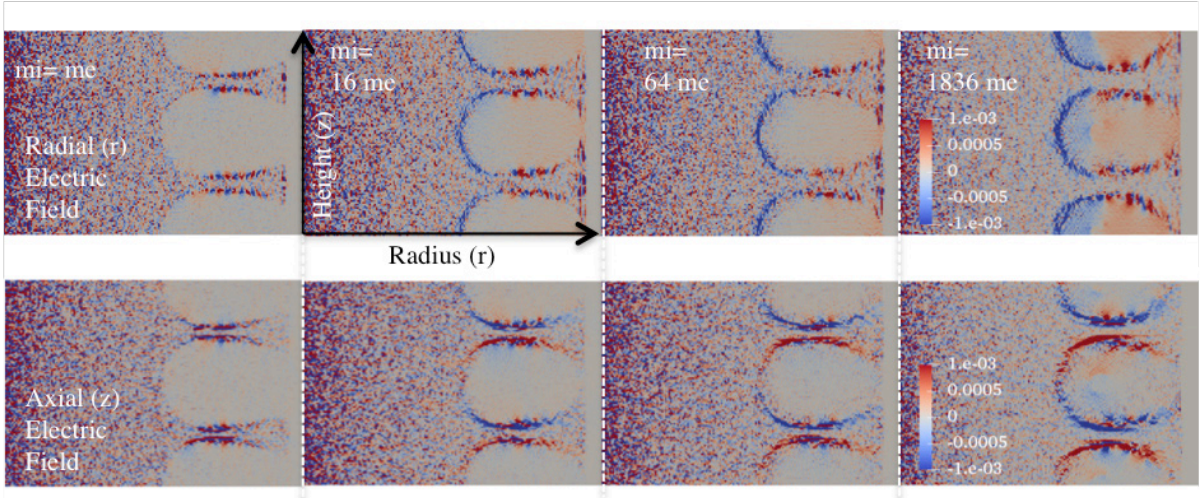

Figure 4. Steady-state equilibrium profiles of electric field as a function of ion mass. Equilibrium profiles show electric field in the radial direction (top row) and axial direction (bottom row) for the same ion masses from Runs 4, 6, 7 and 8 as in Fig. 3.

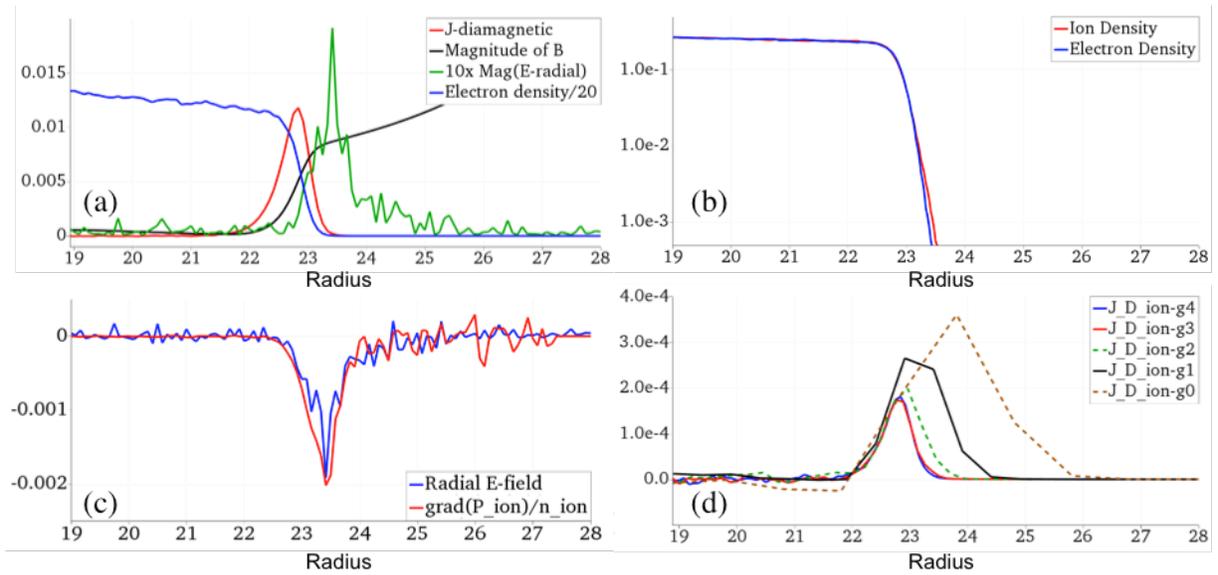

Figure 5. Radial profiles along Line 1 during steady-state equilibrium: a) Magnitude of B-field and E-field multiplied by 10, Diamagnetic current density and Electron density divided by 20, b) Electron and Ion density, c) Radial electric fields and gradient of ion pressure divided by ion density, and d) Ion diamagnetic current density as a function of grid resolution. For plots (a), (b) and (c), the results are from Run 1 with the grid resolution of 540x360. For plot (d), grid resolutions of g0 (45x30). g1 (90x60), g2 (180x120), g3 (360x240) and g4 (540x360) are used for numerical convergence investigation from Runs 1, 9, 10, 11 and 12 from Table ST2.

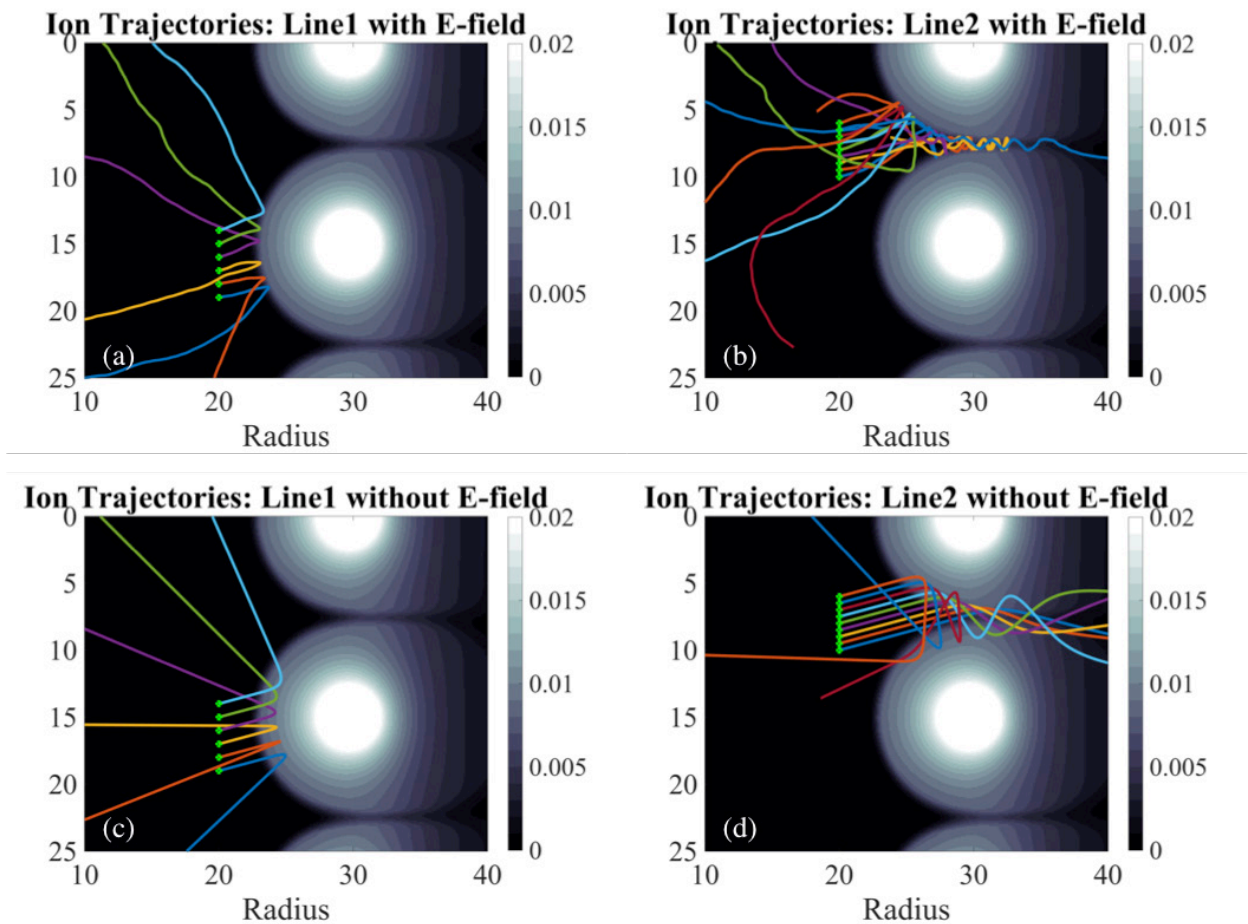

Figure 6. Ion trajectories in the steady-state equilibrium with and without an electric field from Run 1. Fig. 6(a) and 6(b) show ion trajectories with the self-consistent electric field and magnetic field from the simulation. Fig.6 (c) and Fig. 6(d) show ion trajectories calculated with only the magnetic field to highlight the role of the electric field at the boundary in determining the thickness of the boundary layer and the plasma flow collimation.

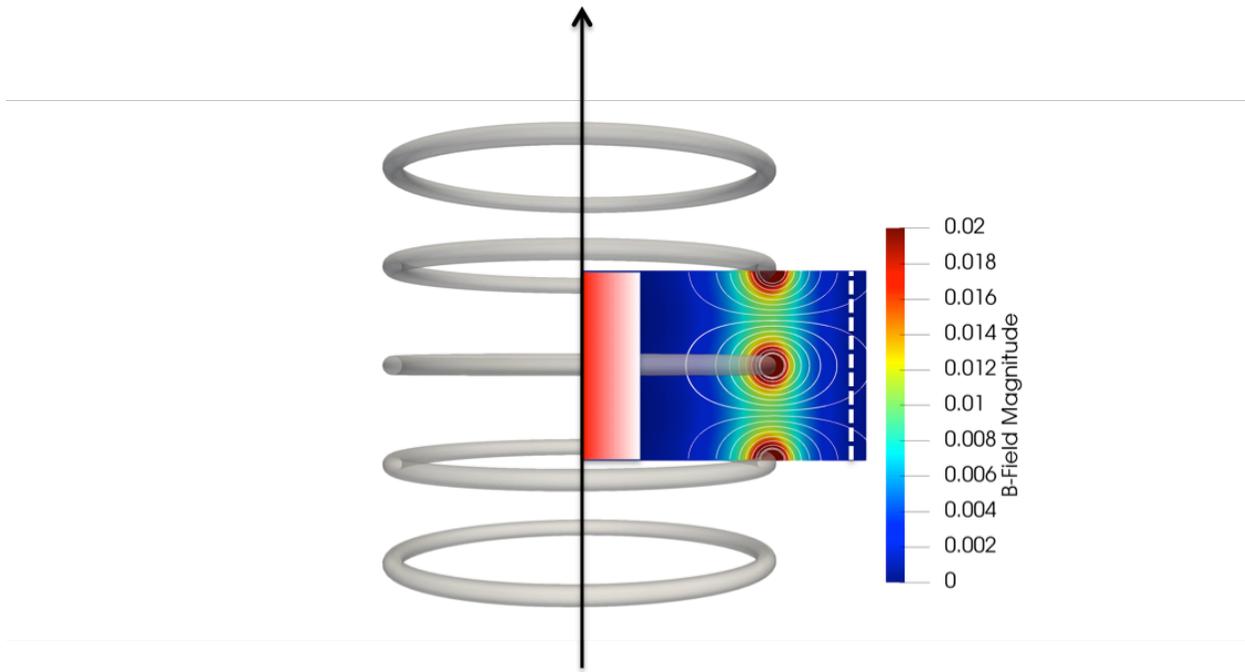

Figure S1. A schematic of a magnetic picket fence plasma system and simulation domain. The schematic shows the contours of magnetic field magnitude and magnetic field lines from the coils without the presence of plasma. The plasma injection region in the central part of the picket fence is shown in graded red and the loss boundary is shown at the right side of the simulation domain as a dotted line.

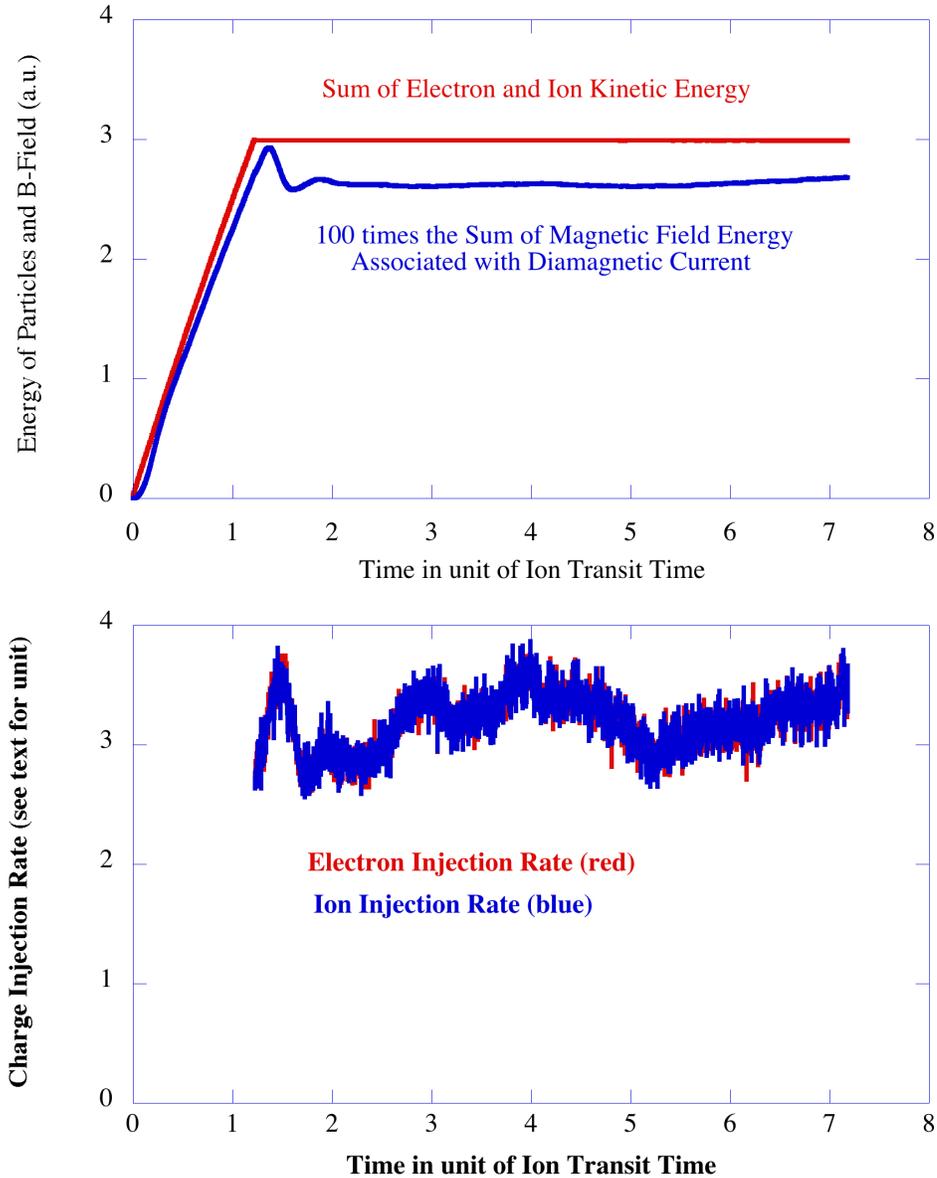

Figure S2. A temporal setup showing the initialization phase and the steady-state phase in the simulation using a unit of ion transit time across the coil diameter of the picket fence. The top row shows the sum of particle kinetic energy in the simulation domain and the sum of magnetic field energy associated with diamagnetic current by the plasma multiplied by 100. The bottom row shows temporal variation of plasma injection rates for electrons and ions to sustain the constant total charge and total particle kinetic energy in the system. The results are from Run 4.

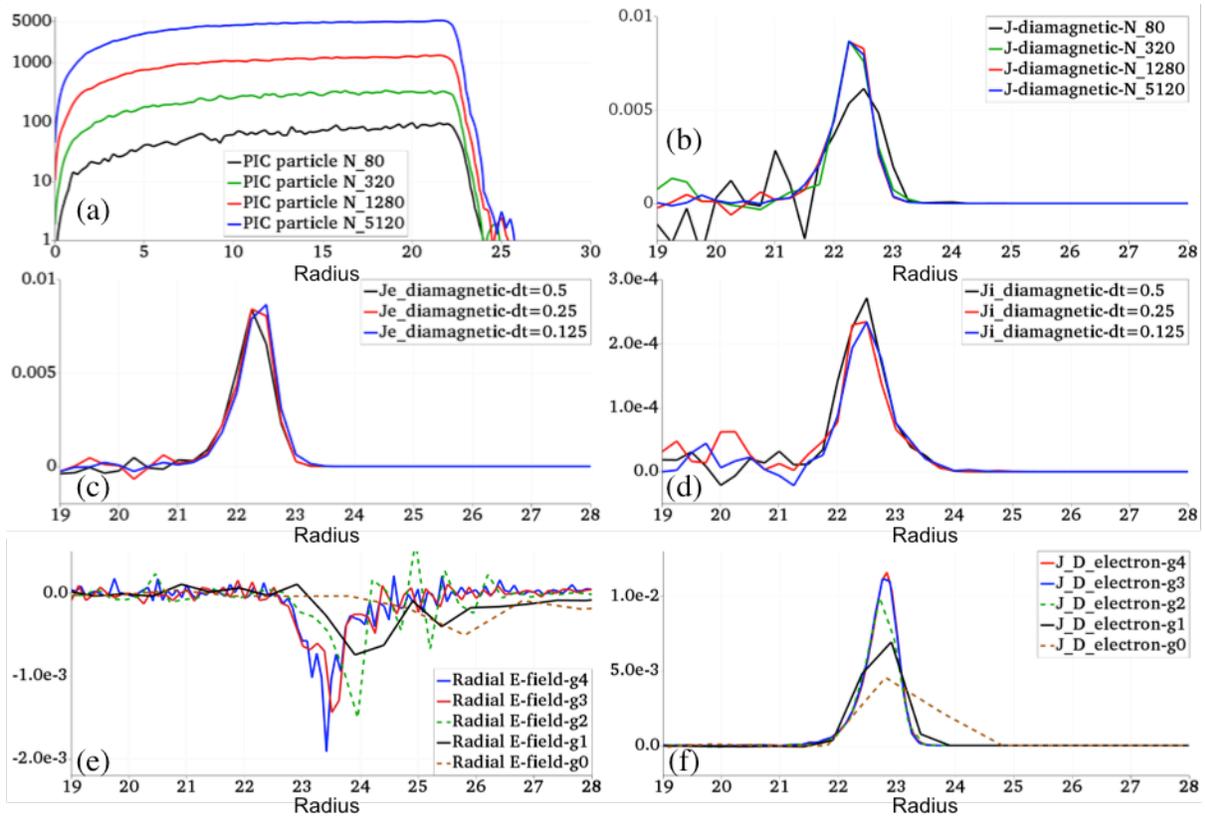

Figure S3. Equilibrium profile results along Line 1 from numerical convergence tests as a function of number of simulation particles (a and b), time step size (c and d) and grid size (e and f). Fig. S3(a) shows the number of simulation particles per each grid and Fig S3 (b) shows the plasma diamagnetic current density at the boundary layer as a function of particles numbers per grid. Figures S3(c) and S3(d) show the electron and ion diamagnetic current density at the boundary layer as a function of time step. Figures S3(e) and S3(f) show the radial electric field and the electron diamagnetic current density as a function of grid size.

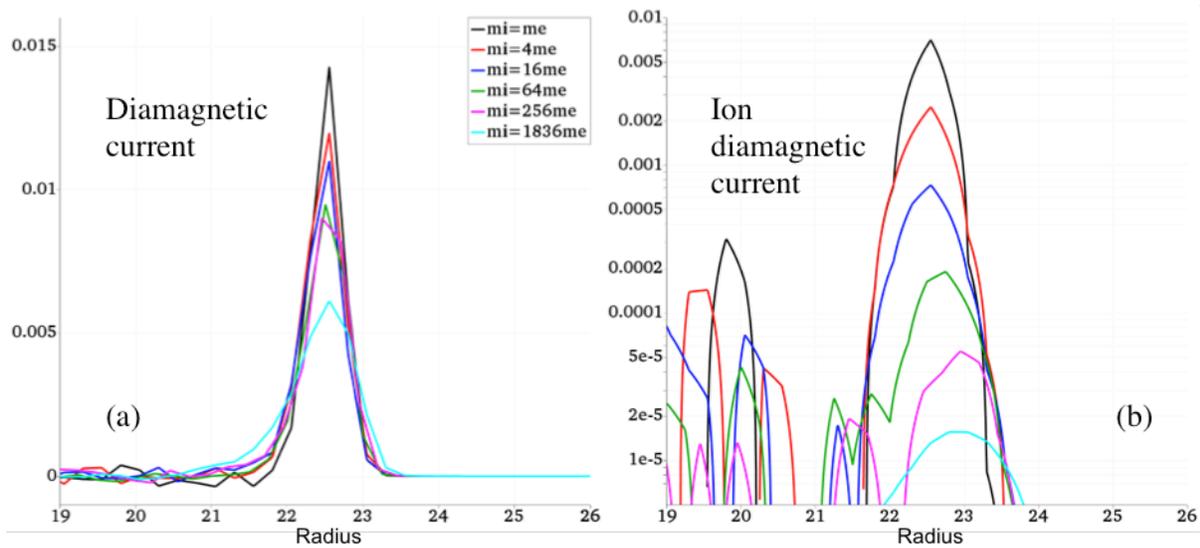

Figure S4. Addendum to Fig. 3 including additional ion mass results. Fig. S4(a) and Fig. S4(b) show the radial profiles of diamagnetic current density and ion diamagnetic current density at the boundary along Line 1 for six different ion masses from Runs 4, 6, 7, 8, 19 and 20 in Table ST2.

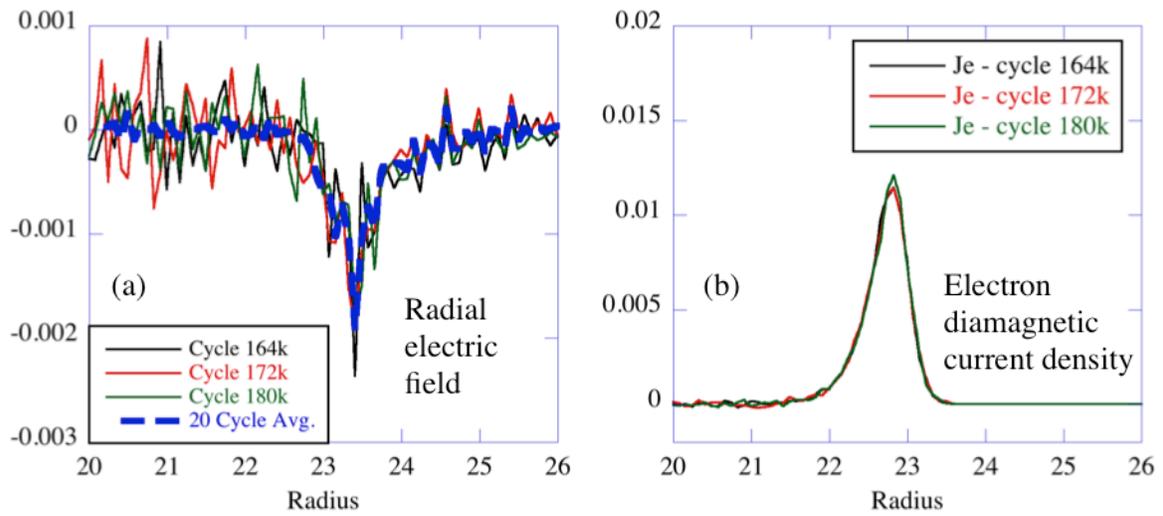

Figure S5. Comparison of noise in simulation results for the electric field and electron diamagnetic current density from Run 1.

| Variables | Run 1 | Run 8 | Magnetic fusion | unit | Magneto-pause | unit |
|---|---|---|---|---|---|---|
| Domain size | 45x30 | 45x30 | 14.7x9.8 | cm | 202x134 | km |
| Grid size | 0.083 | 0.25 | 0.082 | cm | 0.82 | km |
| Coil Diameter | 60 | 60 | 19.8 | cm | 197.6 | km |
| Coli Spacing | 15 | 15 | 4.9 | cm | 49.4 | km |
| Time step | 0.125 | 0.25 | 2.75E-12 | s | 3.23E-05 | s |
| Electron mass | 1.56E-02 | 0.0156 | 9.1E-31 | kg | 9.1E-31 | kg |
| Ion mass | 1 | 28.69 | 1.67E-27 | kg | 1.67E-27 | kg |
| Electron thermal speed | 7.35E-02 | 7.35E-02 | 2.20E+09 | cm/s | 3.75E+08 | cm/s |
| Ion thermal speed | 9.19E-03 | 1.71E-03 | 5.14E+07 | cm/s | 8.76E+06 | cm/s |
| Electron & Ion kinetic energy | 8.44E-05 | 8.44E-05 | 2.77E+03 | eV | 80 | eV |
| Mean value at the boundary along Line 1 | | | | | | |
| Current layer thickness | 0.6 | 0.9 | 0.297 | cm | 2.96 | km |
| B-field | 4.1E-03 | 4.0E-03 | 1325 | Gauss | 22.52 | nT |
| Plasma Density | 0.12 | 0.120 | 2.0E+13 | /cc | 20.0 | /cc |
| Electron gyroradius | 0.28 | 0.29 | 0.094 | cm | 0.95 | km |
| Ion gyroradius | 2.24 | 12.30 | 4.05 | cm | 40.5 | km |
| Debye length | 0.027 | 0.027 | 8.74E-03 | cm | 0.015 | km |
| Ion inertial length | 2.89 | 15.46 | 5.10 | cm | 51.0 | km |
| Electron transit time | 8.16E+02 | 8.16E+02 | 8.97E-09 | s | 5.27E-02 | s |
| Ion transit time | 6.53E+03 | 3.50E+04 | 3.84E-07 | s | 2.26E+00 | s |

Table ST1. Summary of simulation parameters and results. Table ST1 compares the various simulation parameters and results from Run 1 and Run 8 using normalized code unit (NCU) and simulation parameters and results of Run 8 converted for two plasma regimes relevant for a magnetic fusion system and the Earth's magnetopause.

| Run ID | Case | Vol. Avg. Pressure | Ion Mass (in me) | Grid Size | Time Step | Number of Particles |
|---|---|---|---|---|---|---|
| 1 | Baseline | 5.2E-05 | 64 | g4: 540x360 | 0.25 | 172 M |
| 2 | Pr-0.12 | 1.2E-06 | 64 | g2: 180x120 | 0.25 | 11 M |
| 3 | Pr-0.77 | 7.7E-06 | 64 | g2: 180x120 | 0.25 | 11 M |
| 4 | Pr-5.2 | 5.2E-05 | 64 | g2: 180x120 | 0.25 | 11 M |
| 5 | Pr-7.3 | 7.3E-05 | 64 | g2: 180x120 | 0.25 | 18 M |
| 6 | Mass-1 | 5.2E-05 | 1 | g2: 180x120 | 0.25 | 11 M |
| 7 | Mass-16 | 5.3E-05 | 16 | g2: 180x120 | 0.25 | 11 M |
| 8 | Mass-1836 | 5.0E-05 | 1836 | g2: 180x120 | 0.25 | 12 M |
| 9 | g0 | 5.2E-05 | 64 | g0: 45x30 | 0.25 | 0.67M |
| 10 | g1 | 5.2E-05 | 64 | g1: 90x60 | 0.25 | 2.7 M |
| 11 | g2 | 5.2E-05 | 64 | g2: 180x120 | 0.25 | 11 M |
| 12 | g3 | 5.2E-05 | 64 | g3: 360x240 | 0.25 | 76 M |
| 13 | N80 | 5.2E-05 | 64 | g2: 180x120 | 0.25 | 0.73 M |
| 14 | N320 | 5.2E-05 | 64 | g2: 180x120 | 0.25 | 2.9 M |
| 15 | N1280 | 5.2E-05 | 64 | g2: 180x120 | 0.25 | 11 M |
| 16 | N5120 | 5.2E-05 | 64 | g2: 180x120 | 0.25 | 47 M |
| 17 | dt-0.125 | 5.2E-05 | 64 | g2: 180x120 | 0.125 | 11 M |
| 18 | dr-0.5 | 5.2E-05 | 64 | g2: 180x120 | 0.5 | 11 M |
| 19 | Mass-4 | 5.2E-05 | 4 | g2: 180x120 | 0.25 | 11 M |
| 20 | Mass-256 | 5.0E-05 | 256 | g2: 180x120 | 0.25 | 11 M |

Table ST2. Summary of simulation runs with key parameters. Results from Run 1 through 12 are discussed in the main text and the results from Run 13-20 are discussed in the supplemental materials.